\documentstyle[aps,prl,twocolumn,floats,psfig]{revtex}

\begin{document}

\bibliographystyle{prsty}

\title{ 
%
Spin tunneling via dislocations in Mn$_{12}$ acetate crystals
\vspace{-1mm}
}

\author{
E. M. Chudnovsky and  D. A. Garanin
}

\address{
Department of Physics and Astronomy, Lehman College, City University of New York, \\
250 Bedford Park Boulevard West, Bronx, New York 10468-1589 \\
\smallskip
{\rm(Received 26 April 2001)}
\bigskip\\
\parbox{14.2cm}
{\rm
We show that dislocations should be the main source of spin tunneling in Mn$_{12}$ crystals.
Long-range strains caused by dislocations produce broad distribution of relaxation times that has been seen in many experiments.
When the external magnetic field is applied along the c-axis of the crystal, local rotations of the magnetic anisotropy axis due to dislocations result in the effective local transverse magnetic field that unfreezes odd tunneling resonances.
Scaling law is derived that provides universal description of spin tunneling for all resonances.
\smallskip
\begin{flushleft}
\end{flushleft}
} 
} 
\maketitle

Mn$_{12}$ acetate crystals exibit quantum magnetic phenomena at a macroscopic scale.
They have centered tetragonal structure with $a=17.319 \AA$ and $c=12.388 \AA$ as lattice parameters\cite{lis80,henetal97}.
Spin-10 Mn$_{12}$ molecules at the sites of the lattice show magnetic bistablity due to the 65K barrier between the spin-up and spin-down states\cite{sesgatcannov93}.
Quantization of spin levels manifests itself in a spectacular stepwise magnetic hysteresis \cite{frisartejzio96}.
Other observations include memory effects \cite{wersesgat99}, non-exponential relaxation  \cite{wersesgat99,bokkenwal00}, and a peculiar crossover between thermal and quantum behavior \cite{kenetal00,bokkenwal00}.
Theoretical models explain tunneling  in Mn$_{12}$ by phonons \cite{garchu97,luibarfer98,leulos99eplprb,chugar00eplleulos00epl}, nuclear spins \cite{prosta98,garchusch00}, dipolar fields \cite{prosta98,chu00prlprosta00werpauses00}, fourth-order magnetic anisotropy \cite{harpolvil96}, Landau-Zener effect \cite{dobzve97,leulos00zen}, Jahn-Teller effect \cite{garg98}, etc.
Explanation of some observations remains controversial though.

Firstly, there is no agreement among researchers on what causes tunneling in Mn$_{12}$. 
The forth-order transverse anisotropy cannot account for odd resonances.
Besides, it is too weak, as are hyperfine and dipolar fields, to provide the actual tunneling rate.
Secondly, in all samples studied to date, a ``minor species" of Mn$_{12}$ has emerged that exhibits faster magnetic relaxation than the ``major species" \cite{Wern2ndSpecies}.
%
%
Finally,  the time dependence of the magnetic relaxation  in Mn$_{12}$ has not been understood. 
If all Mn$_{12}$ molecules were subject to the same crystal field, the magnetic relaxation would be strictly exponential in time. 
In reality, however, the relaxation deviates from the exponential in the kelvin range and is clearly non-exponential in the subkelvin range \cite{wersesgat99,bokkenwal00}.

\begin{figure}[t]
\unitlength1cm
\begin{picture}(11,6.3)
\centerline{\psfig{file=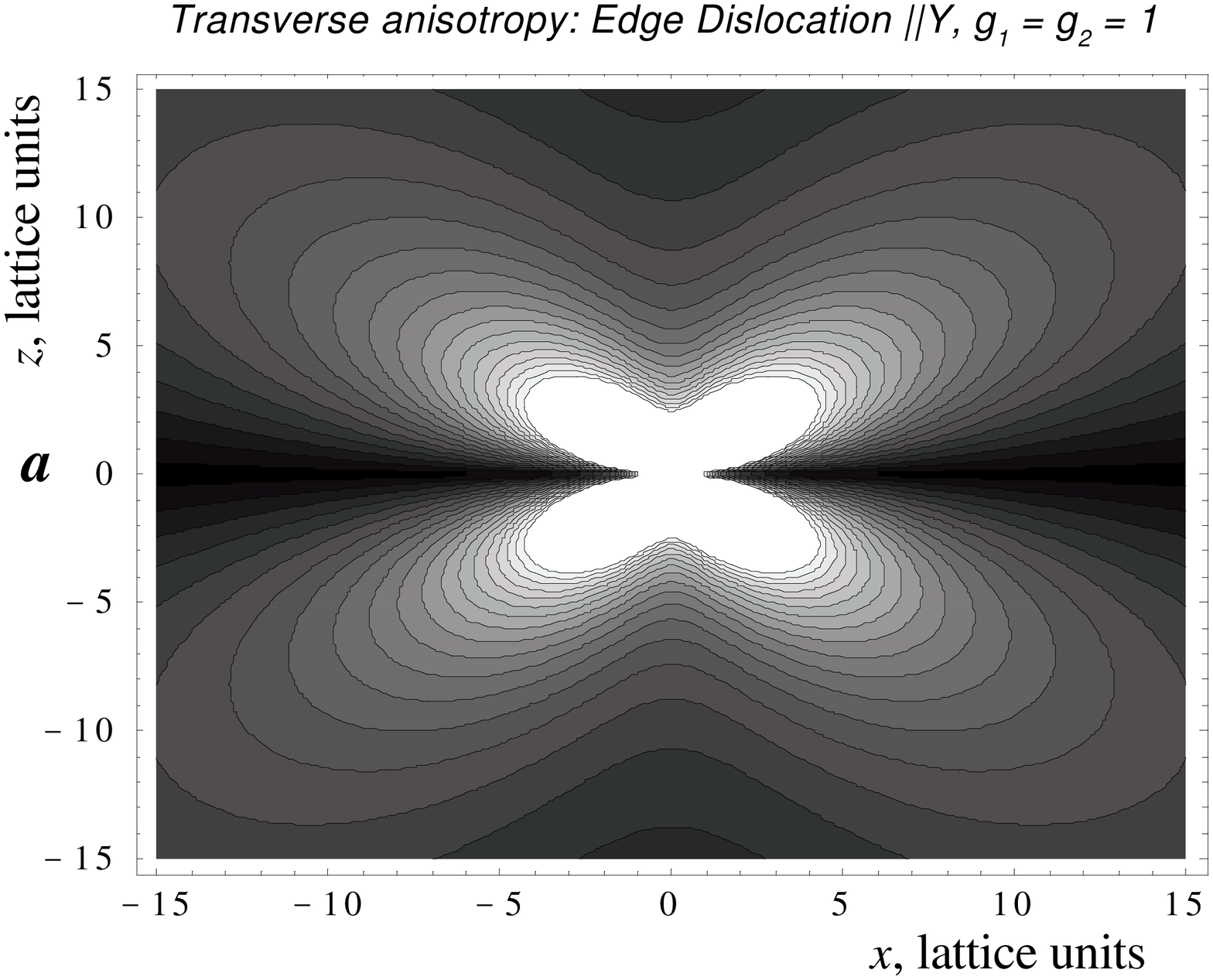,angle=-90,width=9cm}}
\end{picture}
\begin{picture}(11,5.8)
\centerline{\psfig{file=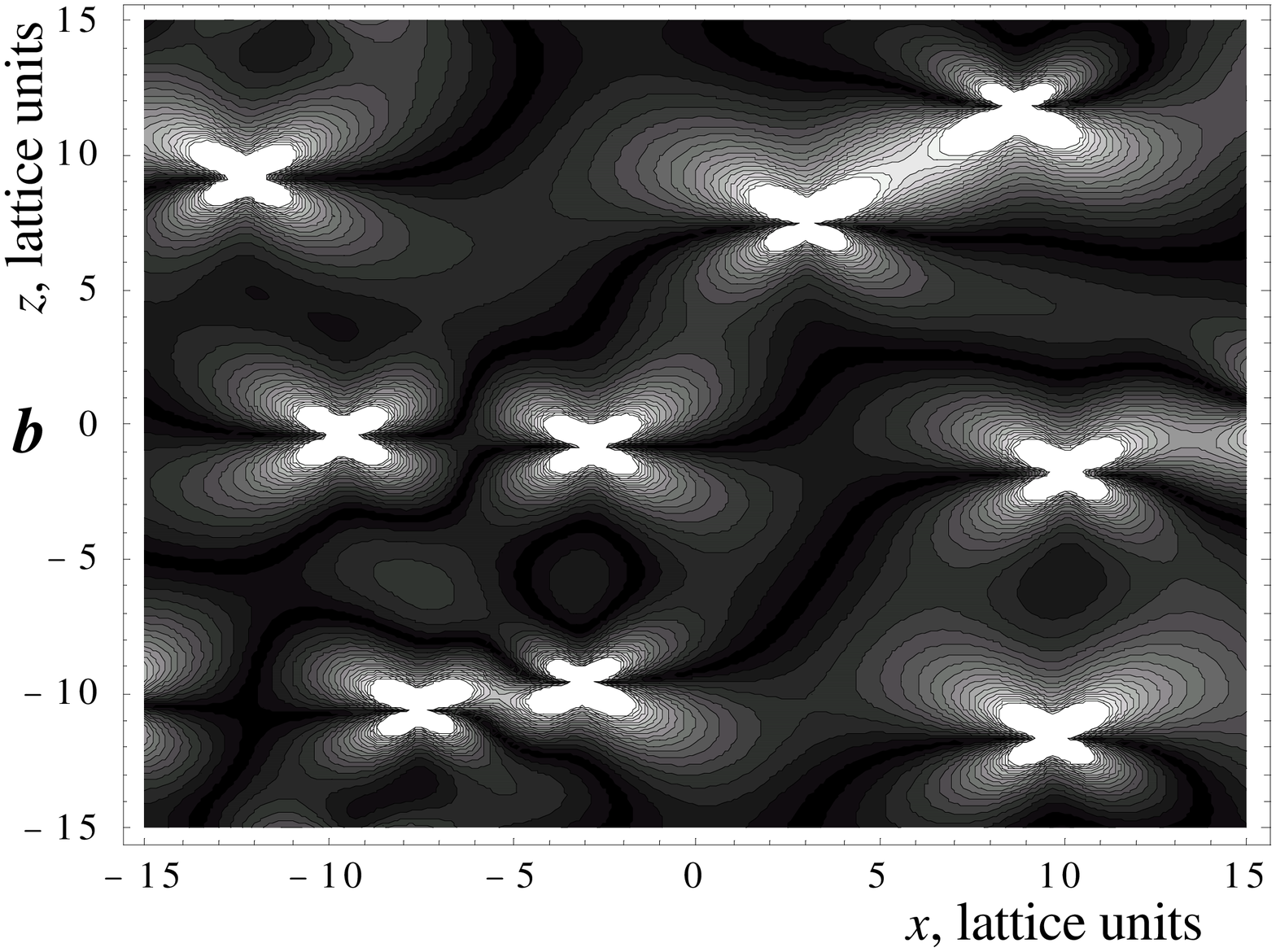,angle=-90,width=9cm}}
\end{picture}
\caption{ \label{fig_strains} 
Contour plot of transverse anisotropy $E$ of Eq.\ (\protect\ref{hamRot}) created by: $a$ - one edge dislocation along the $y$-axis; $b$ - randomized array of  dislocations.
(Grey scales are arbitrary.)
}
\end{figure}

In this Letter we show that dislocations, undoubtedly present in Mn$_{12}$ crystals, provide long range deformations (see Fig.\ \ref{fig_strains}) which should be the main source of spin tunneling. 
The  ``minor" and ``major" species differ on the strength of the deformation, that is, on their distance from the dislocation cores, with no sharp boundary between the two.
Local rotations of the anisotropy axes  due to dislocations are responsible for odd tunneling resonances.
Broad distribution of deformations causes broad distribution of tunneling rates.
We compute the relaxation law in a crystal with dislocations and show  that it obeys scaling that can be tested in experiment. 

We study Hamiltonian
\begin{equation}\label{ham}
{\cal H} = - D S_z^2 - H_z S_z + {\cal H}_{\rm me},
\end{equation}
where $S_z$ is the $z$-component of the spin operator, $S=10$,  $D=0.65$~K, $H_z$ is the magnetic field applied along the $z$-axis ($c$-axis of the crystal), and ${\cal H}_{\rm me}$ is the magnetoelastic coupling.
The Hamiltonian of Mn$_{12}$  also contains crystal fields of fourth order on the spin operator, magnetic dipole interactions, and hyperfine interactions.
We neglect them in order to emphasize the effect of dislocations.
The magnetoelastic coupling in Mn$_{12}$ is of the form \cite{harpolvil96,leulos99eplprb} 
\begin{eqnarray}\label{hamme}
&&
{\cal H}_{\rm me} = g_1 D ( \varepsilon_{xx} - \varepsilon_{yy} ) (S_x^2-S_y^2)
+ g_2 D \varepsilon_{xy} \{ S_x, S_y \}
\nonumber\\
&& \qquad
{} +g_3 D ( \varepsilon_{xz} \{ S_x, S_z \} + \varepsilon_{yz} \{ S_y, S_z \} )
\nonumber\\
&& \qquad
{} +g_4 D ( \omega_{xz} \{ S_x, S_z \} + \omega_{yz} \{ S_y, S_z \} ) ,
\end{eqnarray}
where
\begin{equation}\label{epsDef}
\epsilon_{{\alpha}{\beta}}=\frac{1}{2}\left(\frac{\partial u_\alpha}
{\partial x_ \beta} + \frac{ \partial u_ \beta }{\partial x_\alpha} \right),
\quad 
\omega_{ \alpha \beta }=\frac{1}{2}\left(\frac{ \partial u_ \alpha }
{ \partial x_ \beta } - \frac{ \partial u_ \beta }{ \partial x_ \alpha }  \right)
\end{equation}
are linear deformation tensors, ${\bf u}$ being the displacement.

The full classification of relevant elastic deformations due to different types of dislocations will be done in a longer  paper.
In this Letter, we will illustrate the effect of static deformations on tunneling by considering edge dislocations running perpendicular to the anisotropy direction along the $y$-axis. 
For the YZ extra crystallographic plane inserted at $z>0$ one obtains (see, e.g., Ref.\ \cite{lanlif7}): $\varepsilon_{xy} = \varepsilon_{yz}  =  \omega_{yz} = 0$,
\begin{eqnarray}\label{epsEdge}
&&
\varepsilon_{xz} = \frac{a}{2\pi} \frac x {  x^2 + z^2 },
\quad 
\omega_{xz} = \frac{a}{4\pi}  \frac {x (x^2 - z^2)} {  (1-\sigma) (x^2 + z^2)^2 } 
\nonumber\\
&&
 \varepsilon_{xx} - \varepsilon_{yy}  =
\frac{a}{4\pi} z \frac { (2\sigma-3)(x^2 + z^2) + 2 z^2 }{ (1-\sigma) (x^2 + z^2)^2 },
\end{eqnarray}
where $0<\sigma <1/2$ is the Poisson elastic coefficient (we will use $\sigma=0.25$).
If the extra plane is inserted at $z<0$, the above expressions change their sign.

Due to the terms of the type $S_x S_z$, etc., in  the Hamiltonian, the local easy axis deviates from the $z$-direction.
In a locally rotated coordinate system ($x',y',z'$) that restores the normal form of the crystal field, saving terms linear on deformation, one obtains 
\begin{equation}\label{hamRot}
{\cal H} = - D S_{z'}^2 - H_z S_{z'} + E  (S_{x'}^2 - S_{y'}^2)  - H_{x'} S_{x'} ,
\end{equation}
where 
\begin{equation}\label{EHxDef}
E = g_1 D ( \varepsilon_{xx} - \varepsilon_{yy} ), 
\quad
H_{x'} = \frac 12 (g_3  \varepsilon_{xz} + g_4  \omega_{xz})  H_z.
\end{equation}
A few observations are in order.
Dislocation generates the transverse anisotropy  of strength $E$ that decays as $1/r$ on the distance from the dislocation core.
Due to the slow decay of $E$, a single dislocation produces spin tunneling at a large number of crystal sites. 
The field  applied along the $c$-axis of the crystal generates the transverse field due to the local rotation of the easy axis by the dislocation.
Above the XY plane (at $z>0$) the transverse field is directed along the hard axis, while at $z<0$ the transverse field is along the medium axis. 

For numerical work we choose $g_1=g_2=g_3=g_4=1$ (see, e.g., Refs.\  \cite{chugar00eplleulos00epl} and references therein).
The spatial dependence of the transverse anisotropy $E$ due to a single edge dislocation, described by  Eqs.\ (\ref{epsEdge}) and  (\ref{EHxDef}),  is shown in Fig.\ \ref{fig_strains}a.
The effect of many dislocations is additive as long as the linear elastic theory is applied, which is correct outside dislocation cores.
Fig.\ \ref{fig_strains}b shows a typical pattern of the transverse anisotropy produced by an array of edge dislocations, obtained by randomization of the quadratic lattice of dislocations with alternating orientation of the extra plane. 

\begin{figure}[t]
\unitlength1cm
\begin{picture}(11,6.5)
\centerline{\psfig{file=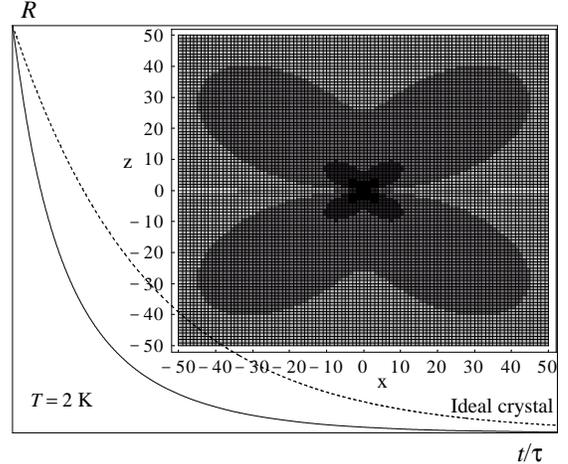,angle=-90,width=9cm}}
\end{picture}
\caption{ \label{fig_activ} 
Relaxation curves for the $100 \times 100$  Mn$_{12}$ lattice around a single edge dislocation along the $y$-axis at $T=2$~K and $H=0$.
Faster relaxation of  Mn$_{12}$ molecules caused by the dislocation leads to a nonexponential form of $R(t/\tau)$, where $\tau$ is the relaxation  time for an ideal sample.
Dashed line: Pure thermal exponential in the ideal Mn$_{12}$ crystal.
Inset: Tunneling levels $m_{b}=0,-1,-2,-3,-4$ at different points of the lattice  ($m_{b}=0$ in the small lightmost regions corresponds to the 65K barrier in an ideal crystal).
}
\end{figure}

Due to the dislocations the local energy barrier between spin-up and spin-down states is lower than in the ideal crystal.
It can be calculated perturbatively from Eq.\ (\ref{hamRot}).  
In addition to the barrier reduction for a classical spin, the
barrier in the quantum case is further lowered at the discrete values of $H_z$, 
\begin{equation}\label{ResCond}
H_{zk} = k \sqrt{D^2-E^2}, \qquad k=0,\pm 1, \ldots, \pm 2S
\end{equation}
that provide resonant tunneling between matching spin levels $m$ and $m'=-m-k$ \cite{frisartejzio96,garchu97,luibarfer98,kecgar01}. 
At $H_z=H_{zk}$ the effective height of the barrier,
\begin{equation}\label{BarEff}
U^{\rm eff}=D(S^2-m_b^2) - H_z(S+m_b),
\end{equation}
is determined by $m=m_b<0$ that corresponds to the lowest pair of levels, $m$ and $m'$, whose tunneling splitting, $\Delta_{mm'}$,  is greater than the sum of their widths, $\Gamma_{mm'}=\Gamma_m+\Gamma_{m'}$. 
Eq.\ (\ref{BarEff}), even though it accounts for only one tunneling pair of levels,  is a good approximation for the relaxation problem (see discussion after Eq.\ (5.15) and Fig.\ 5 of Ref.\ \cite{garchu97}).
The splitting is determined by high powers of $E$ and $H_{x'}$, and, thus, strongly depends on coordinates [see Eq.\ (2.6) of Ref.\ \cite{garchu97} for the effect of the transverse field and Eq.\ (4) of Ref.\ \cite{garchu99} for the effect of transverse anisotropy]. 
Consequently, the dependence of $m_b$ on $\Gamma_{mm'}$ is weak.  
In the kelvin range we will use the experimental value $\Gamma_{mm'}\simeq 200$~Oe \cite{frisarzio98}.

To illustrate the barrier reduction by dislocations in the thermally activated regime (kelvin range), we numerically diagonalized Eq.\ (\ref{hamRot}) for the $100 \times 100$  Mn$_{12}$ lattice around a single edge dislocation at $T=2$~K and $H=0$.
Using the criterion described above, we determine tunneling levels $m_b$ and the effective barriers $U_i^{\rm eff}$ from Eq.\ (\ref{BarEff}) for all sites $i$ of the pattern and construct the relaxation curve
\begin{equation}\label{RelCurve}
R(t) = \frac 1 N \sum_i \exp(-\alpha_i t /\tau), 
\end{equation}
where $\alpha_i = e^{ ( U_\infty - U_i^{\rm eff} )/T}$, $U_\infty \equiv DS^2$ is the unperturbed barrier, and $\tau=\tau_0 e^{ U_\infty/T}$ is the relaxation time far from the dislocation.
One can see from Fig.\ \ref{fig_activ} that the effect of dislocations on relaxation is profound.
Even for one dislocation per $100 \times 100$ sites in the XZ plane, which corresponds to the concentration of dislocations as small as $c=10^{-4}$, more than half of Mn$_{12}$ molecules relax faster than in the ideal crystal. 
The list of energy barriers  reads 65K (64), 64.35K (4700), 62.38K (4944), 59.06K (244), 54.31K (36), 48.32K (8), 30.73K (4), with the number of corresponding crystal sites shown in the brackets.

\begin{figure}[t]
\unitlength1cm
\begin{picture}(11,6.5)
\centerline{\psfig{file=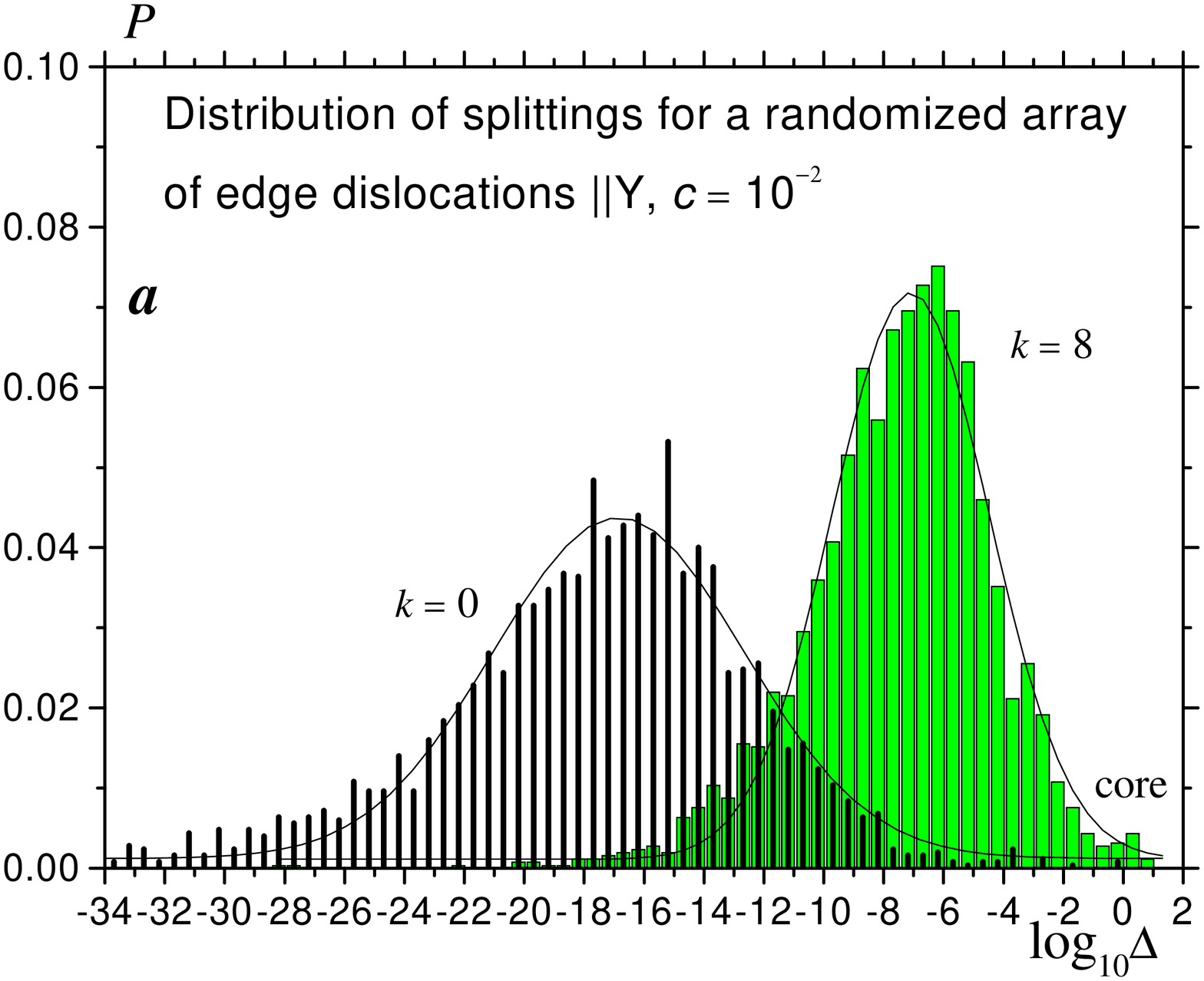,angle=-90,width=9cm}}
\end{picture}
\begin{picture}(11,6.3)
\centerline{\psfig{file=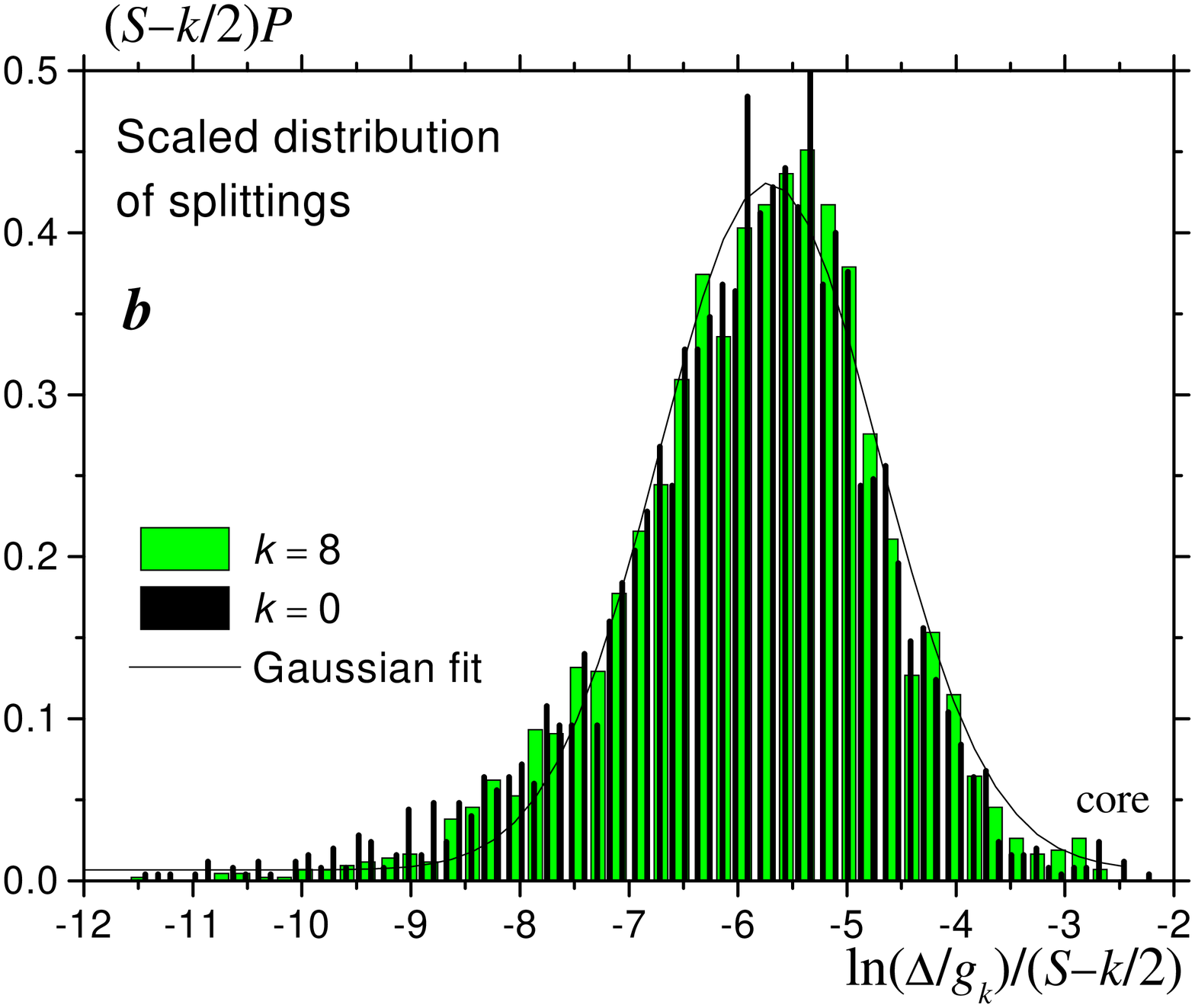,angle=-90,width=9cm}}
\end{picture}
\caption{ \label{fig_distr} 
$a$ - Distribution of level splittings (in kelvin) created by a randomized array of edge dislocations with $c=10^{-2}$ for resonances with $k=0$ and $k=8$; 
$b$ - scaled representation with $g_k$ of Eq.\ (\protect\ref{SplitPert}). }
\end{figure}

At temperatures below 1K, where spin relaxation occurs via tunneling from the ground state, the influence of dislocations becomes even more dramatic. 
In contrast to the small barrier reduction in the thermally activated regime,  here the role of dislocations is to provide the main source of spin relaxation.
The resulting level splittings are distributed over many decades.
Fig.\ \ref{fig_distr}a shows distributions of the ground-state level splittings obtained numerically for a randomized array of edge dislocations with concentration $c=10^{-2}$ for $k=0$ and $k=8$ resonances.
Since outside dislocation cores the perturbations of the uniaxial Hamiltonian are small, one can scale the splitting distributions for even values of $k$ using the perturbative Eq.\ (4) of Ref.\ \cite{garchu99} for the ground-state splitting ($m=-S$, $m'=-m-k=S-k$),
\begin{eqnarray}\label{SplitPert}
&&
\Delta_{ki} = g_k \left( \frac {E_i} {8D} \right)^{S-k/2}
\nonumber\\
&&
g_k = \frac {8D} { [ (S-k/2-1)!]^2 } \sqrt{ \frac { (2S-k)! (2S)!  } {k!} } .
\end{eqnarray}
Then the distribution of $\ln(\Delta_{ki}/g_k)/(S-k/2)=E_i/(8D)$ does not depend on $k$, see Fig.\ \ref{fig_distr}b.
For $k\neq 0$ there is also a transverse field $H_{x'}$ in addition to the transverse anisotropy $E$ in Eq.\ (\ref{hamRot}), which should modify  Eq.\ (\ref{SplitPert}).
This modification, however, is small away from the dislocation cores.
A more significant effect of the transverse field generated by the dislocations is unfreezing of tunneling resonances with odd values of $k$.
Our calculation shows that the distribution of the level splittings for odd $k$ is similar to that for even $k$ in Fig.\ \ref{fig_distr}. 

\begin{figure}[t]
\unitlength1cm
\begin{picture}(11,6.5)
\centerline{\psfig{file=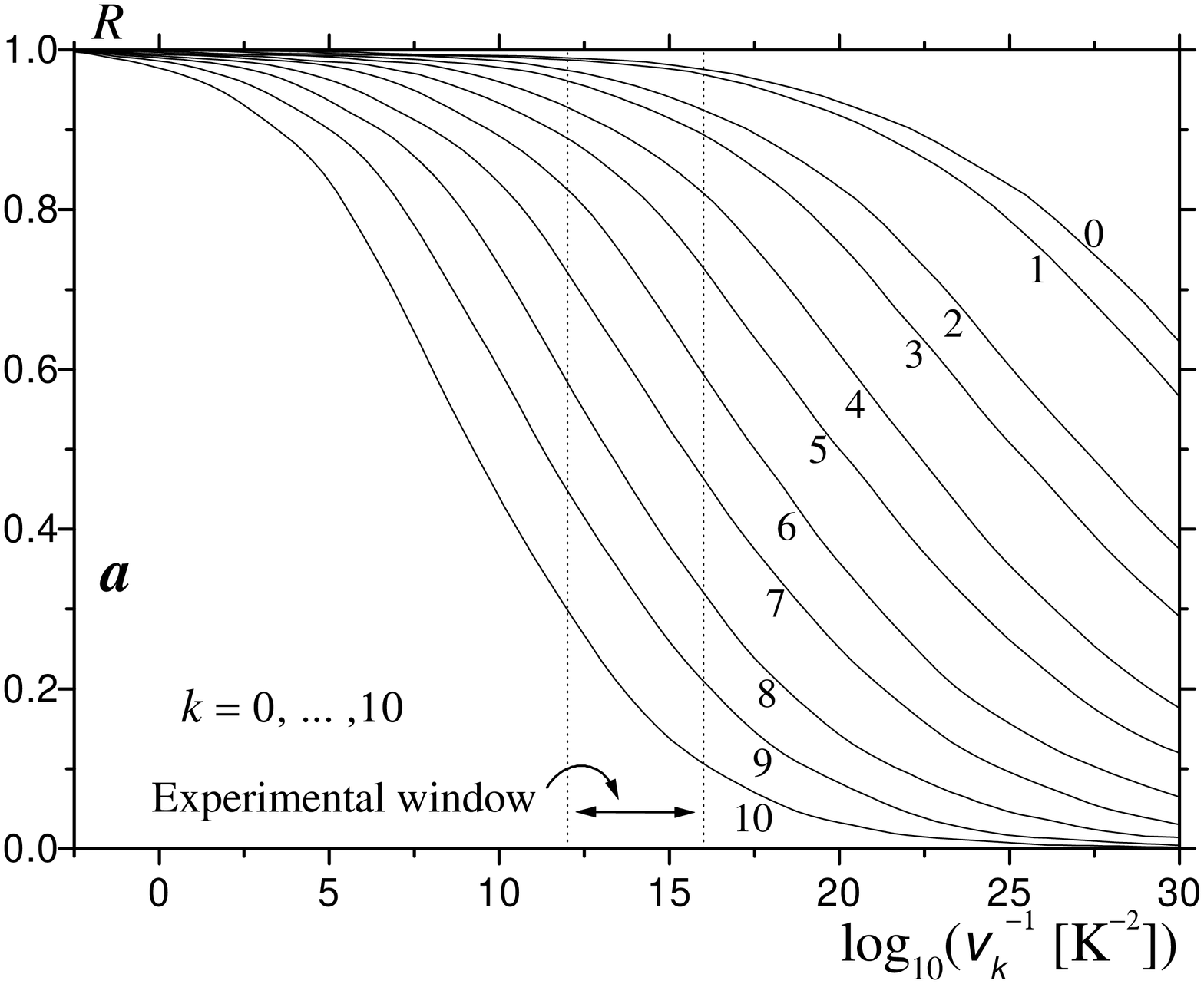,angle=-90,width=9cm}}
\end{picture}
\begin{picture}(11,6.2)
\centerline{\psfig{file=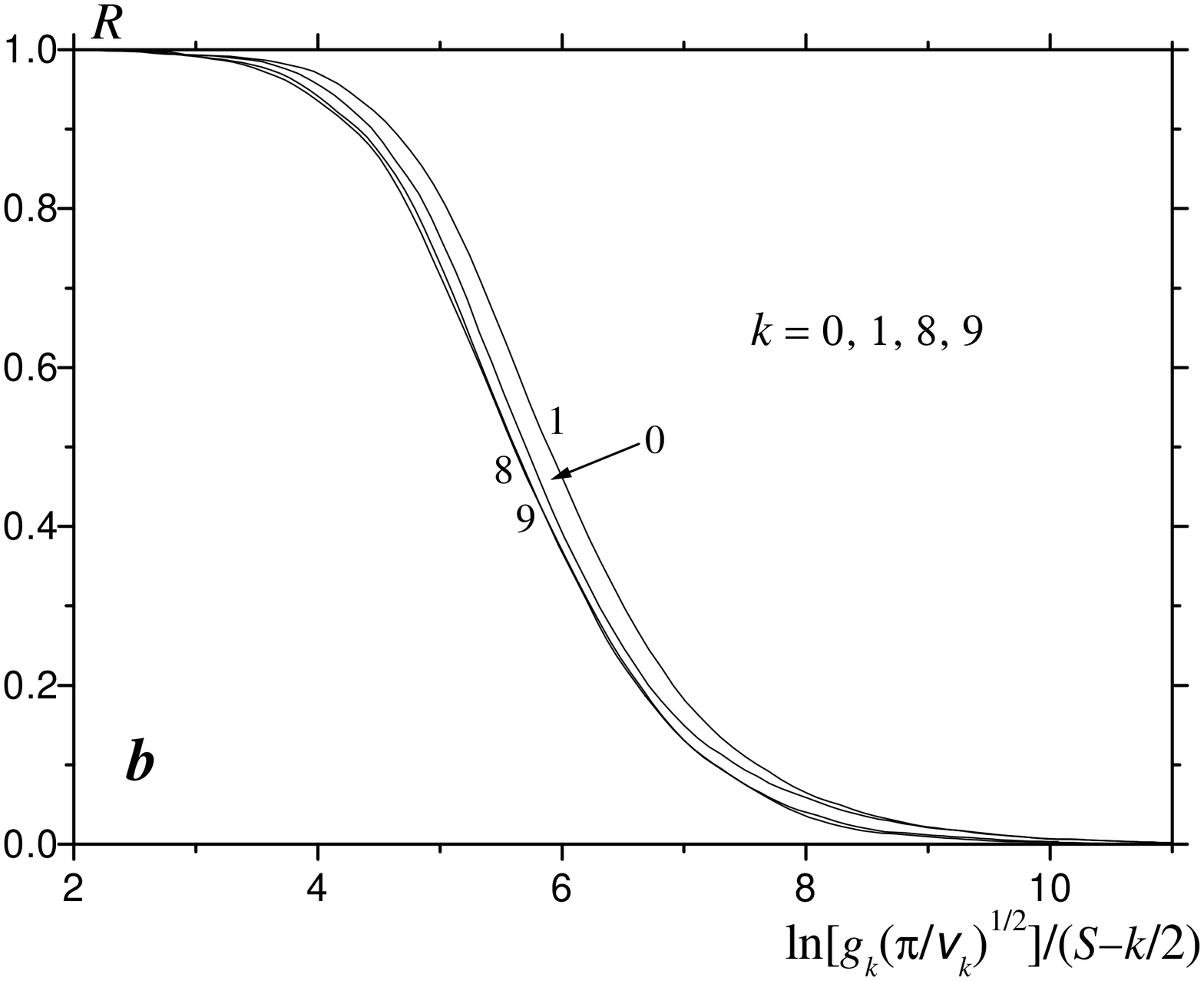,angle=-90,width=9cm}}
\end{picture}
\caption{ \label{fig_rel} 
$a$ - Spin relaxation as a function of the field sweep rate $v_k=(2S-k)dH_z/dt$ for a randomized array of edge dislocations with $c=10^{-2}$ at different tunneling resonances $k$. 
$b$ - scaled representation with $g_k$ of Eq.\ (\protect\ref{SplitPert}). 
$dH_z/dt$ is measured in $K^2\simeq 10^{11}$ T/s.
}
\end{figure}

Let us study now spin relaxation produced by sweeping the field $H_z$ through a tunneling resonance.
The population of the metastable state with $m=-S$ obeys 
\cite{garchu97}
\begin{equation}\label{KinEq}
\dot N_{-S} = - N_{-S} \frac{  \Delta^2 \Gamma } 
{ ( \varepsilon_{-S} -  \varepsilon_{S-k} )^2 +  \Gamma ^2 },
\end{equation}
where $ \varepsilon_{-S} -  \varepsilon_{S-k} = v_k t$ and $v_k\equiv (2S-k) dH_z/dt$.
Integrating this equation one obtains the probability for the molecule to remain in the metastable well after crossing the resonance, $P = \exp(-\pi \Delta^2/v_k)$, which resembles the Landau-Zener formula (see also Ref.\ \cite{leulos00zen}).
The relaxation curve $R(v_k)$ for a sample with dislocations
\begin{equation}\label{RelCurvev}
R(v_k) = \frac 1 N \sum_i \exp(-\pi \Delta_i^2/v_k)
\end{equation}
at different resonances $k$ is shown in Fig.\ \ref{fig_rel}.
At odd $k$ relaxation is produced by the transverse field $H_{x'}$ in combination with the transverse anisotropy.
For $k=1$ the effect of $H_{x'}$ is still small and the corresponding relaxation curve is noticeably shifted to the right.

Due to the wide distribution of splittings, the sweeping rate  needed to make different Mn$_{12}$ molecules to relax stretches over many decades. 
Consequently, $R(v_k)$ can only be plotted on the log scale.
On that scale any individual exponential becomes a step function,  $ \exp(-\pi \Delta_i^2/v_k) \Rightarrow \theta( 1 -\pi \Delta_i^2/v_k )$.
Since $R(v_k)$ represents the fraction of Mn$_{12}$ molecules which have $\pi \Delta_i^2 < v_k $ and, thus, do not relax, $R(v_k)$ and the distribution of splittings are related as the integral and the derivative. 
Correspondingly, $R(v_k)$ for different $k$ can be scaled the same way as in Fig.\ \ref{fig_distr}b.
Deviations from the perfect scaling in Fig.\ \ref{fig_rel}b are due to the transverse field.

The effect of dislocations, even at moderate concentrations, appears to be much stronger than the effect of transverse fields $H_\perp$ from nuclear spins and dipole interactions.
Simple arguments of the perturbation theory show that the effect of dislocations becomes comparable with that of $H_\perp$ at $E \sim H_\perp^2/(DS^2)$.
With the help of Eqs.\ (\ref{epsEdge})  and (\ref{EHxDef}) one finds that for $c > 10^{-6}$ the effect of dislocations on tunneling is greater than the effect of  hyperfine and dipole fields at almost all crystal sites.

We have shown that dislocations in Mn$_{12}$ crystals should be the main source of spin tunneling in the kelvin and subkelvin temperature range. 
Local rotations of the easy axes due to dislocations give rise to the effective transverse field which unfreezes odd resonances.
Tunneling lifetimes due to dislocations are spread over many decades, which  results in the stretched relaxation, especially pronounced in the subkelvin range.
Distribution of the tunneling rates and the relaxation law obey scaling which does not depend on the type of crystal defect. 
That scaling should be seen in  experiment.
Eventually, it may also become possible to observe magnetization patterns on the surface of Mn$_{12}$ crystals, shown in Fig.\ \ref{fig_strains}.

We thank Jonathan Friedman for useful remarks.
This work has been supported by the NSF Grant No. 9978882.

\vspace{-0.5cm}


\end{document}